\documentclass[article,12pt,showpacs,showkeys]{revtex4}
\usepackage{amsfonts,subfigure,graphicx,amsmath,mathtools,bm,multirow}

\begin{document}
\title[Short Title]{Modulation of entanglement between two oscillators separated in space with an optical parametric amplifier}
\author{Cheng-Hua Bai}
\affiliation{Department of Physics, College of Science, Yanbian
University, Yanji, Jilin 133002, People's Republic of China}
\author{Dong-Yang Wang}
\affiliation{Department of Physics, College of Science, Yanbian
University, Yanji, Jilin 133002, People's Republic of China}
\author{Hong-Fu Wang\footnote{E-mail: hfwang@ybu.edu.cn}}
\affiliation{Department of Physics, College of Science, Yanbian
University, Yanji, Jilin 133002, People's Republic of China}
\author{Ai-Dong Zhu}
\affiliation{Department of Physics, College of Science, Yanbian
University, Yanji, Jilin 133002, People's Republic of China}
\author{Shou Zhang}
\affiliation{Department of Physics, College of Science, Yanbian
University, Yanji, Jilin 133002, People's Republic of China}
\begin{abstract}
We propose a scheme to modulate the entanglement between two oscillators separated in space via the squeezing cavity field generated by the optical parametric amplifier instead of injecting the squeezing field directly with the assistance of Coulomb interaction. We show that the Coulomb interaction between the oscillators is the essential reason for the existence of entanglement. Due to the gain of the optical parametric amplifier and the phase of the pump driving the optical parametric amplifier can simultaneously modulate the squeezing cavity field, the radiation pressure interaction between the cavity field and the oscillator is modulated accordingly. We find that there is competing effect between the radiation pressure interaction and the Coulomb interaction for the oscillator which these two interactions act on simultaneously. Therefore, the modulation of entanglement can be achieved with the assistance of Coulomb interaction. The results of numerical simulation show that the present scheme has stronger robustness against the temperature of environment compared with previous schemes in experimentally feasible regimes.
\end{abstract}
\maketitle

\section{Introduction}\label{sec1}
Quantum entanglement~\cite{3531ESPCPS,0981RHPHMHKHRMP}, as a cornerstone of quantum physics, plays a significant role in the foundation of quantum theory and also has potential applications in quantum technology, such as quantum information science~\cite{12JAJDJ} and quantum  metrology~\cite{115VGSLLMNP}. So far, one has had a fairly good understanding of how to generate entanglement among microscopic entities and entanglement has been successfully prepared and manipulated in variously microscopic systems theoretically and experimentally, such as atoms~\cite{1113HFWADZSZKHYNJP,1532WMSSLSZJYLADZHFWSZJOSAB,1490SLSXQSHFWSZPRA,144SLSXQSHFWSZSR,1592SLSQGHFWSZPRA}, photons~\cite{0979HFWSZPRA,0953HFWSZEPJD,1119HFWSZADZXXYKHYOE}, ions~\cite{9574JICPZPRL,02417DKCMDJWNature,14512ACWYCKRBEKDLDJWNature}, Bose-Einstein condensates~\cite{1220LBCPSCHZYJG}, and so on. However, nothing in the principles of quantum mechanics prevents macroscopic systems from attaining entanglement. Recently, there has been considerable interest in investigating entanglement in mesoscopic and even macroscopic systems~\cite{0288SMVGDVPTPRL,0911SHGSANJP,0368JZKPSLBPRA,0572MPADDVOATBAHEPL,0740DVSMPTJPA,08101MJHMBPPRL,1489JQLQQWFNPRA,15110JLBHYZLWEPL}. This is due to the fact that such entanglement might provide explicit evidence for quantum phenomena~\cite{0558KCSMLRPT} and even might possibly help us to clarify the quantum-to-classical transition, as well as the boundary between classical and quantum worlds~\cite{9144WHZPT}. Since mechanical oscillators resemble a prototype of classical systems, they are beginning to be important candidates for the investigation of quantum features at mesoscopic and macroscopic scales. Additionally, with the rapid progress of practical technologies in cavity optomechanics, the mechanical oscillators can be cooled down close to the quantum ground state~\cite{10464ADOMHMARCBMNDSHWMWJWJMMANCNature,11475JDTTDDLJWHMSAWKWLRWSNature,11478JCTPMAAHSNJTHAKSGMAOPNature}. Thus they provide a nature platform to explore quantum entanglement in macroscopic systems.

In recent years, based on the optomechanical systems, some schemes have been brought forward to generate entanglement between macroscopic oscillators from many different angles of view: such as entangling two oscillators in a ring cavity~\cite{0288SMVGDVPTPRL,0911SHGSANJP}, entangling two distantly separated oscillators by utilizing the entangled light fields~\cite{0368JZKPSLBPRA}, entangling two oscillators via a double-cavity set-up by driving squeezing optical fields~\cite{0572MPADDVOATBAHEPL}, entangling a Fabry-P\'{e}rot cavity's two moving mirrors by driving an intense classical laser field~\cite{0740DVSMPTJPA}, entangling two dielectric membranes suspended inside a Fabry-P\'{e}rot cavity~\cite{08101MJHMBPPRL}, entangling two macroscopic mechanical resonators induced by the radiation pressure of a single photon in a two-cavity optomechanical system~\cite{1489JQLQQWFNPRA}, and entangling two movable mirrors in an optomechanical cavity in which a Kerr-down-conversion crystal consisting of a Kerr nonlinear medium and an optical parametric amplifier (OPA) is placed~\cite{15110JLBHYZLWEPL}. In these schemes, however, the previous works~\cite{0572MPADDVOATBAHEPL,0911SHGSANJP} have verified that the injection of the squeezed field is the necessary condition to generate the desired entanglement. Huang and Agarwal~\cite{0911SHGSANJP} proposed a scheme to entangle two separated mechanical oscillators by injecting broad band squeezed vacuum light and laser light into the ring cavity. This scheme showed that the entanglement can be modulated via the squeezing parameter of the input light. In the case of no injection of the squeezed vacuum light, which means that the squeezed vacuum light is replaced by the ordinary vacuum light, there is always no entanglement between the separated oscillators. However, once the incident vacuum light is squeezed, the entanglement exists. Pinard {\it et al.}~\cite{0572MPADDVOATBAHEPL} also proposed a scheme to generate a stationary entangled state of two movable mirrors if and only if the incident fields are squeezed. In Ref.~\cite{15110JLBHYZLWEPL}, even though the entanglement between two mechanical oscillators in an optomechanical cavity can be generated when the injected field is not squeezed, the region of entanglement is discrete and very narrow, so which inevitably brings difficulties to achieve the entanglement in experiment. However, when the injected field is adjusted to the squeezed field, the region of entanglement is continuous and greatly enlarged. In essence, the OPA inside the optomechanical cavity can produce various novel effects including improvement of the cooling of the micromechanical mirror~\cite{0979SHGSAPRA}, affection of the normal-mode splitting behavior of the coupled movable mirror and the cavity field~\cite{0980SHGSAPRA}, achievement of strong mechanical squeezing~\cite{1693GSASHPRA}, and enhancement of the precision of optomechanical position detection~\cite{15115VPHGLSCMFMPRL}. The nonlinear interaction processes between light and OPA have been demonstrated as important sources of squeezed state of the radiation field~\cite{97MOSMSZQO,94DFWGJMQO}. Agarwal and Huang~\cite{1693GSASHPRA} have had the OPA placed inside the optomechanical cavity so that the squeezing cavity field is generated inside the cavity. Via driving the system by the red-detuned laser in the resolved side band limit makes the optomechanical interaction between the movable mirror and the cavity field like a beam-splitter interaction, the state of squeezed photons transfers to phonons with almost 100\% efficiency, the strong mechanical squeezing is thus achieved. Here we propose a scheme to modulate the entanglement between two oscillators separated in space via the squeezing cavity field generated by the OPA instead of directly injecting the squeezing field with the assistance of Coulomb interaction. We show that the Coulomb interaction between the oscillators is the essential reason for the existence of entanglement and there is competing effect between the radiation pressure interaction and the Coulomb interaction for the oscillator which these two interactions act on simultaneously. According to Ref.~\cite{1693GSASHPRA}, the gain of the OPA and the phase of the pump driving the OPA can simultaneously modulate the squeezing cavity field, so the radiation pressure interaction between the cavity field and the oscillator is modulated accordingly. In this way the modulation of the entanglement between the oscillators separated in space is achieved. In addition, we numerically simulate the critical temperature of such entanglement in experimentally feasible regimes. The results show that the laser driving power and the gain of the OPA can improve the critical temperature. Compared with previous schemes~\cite{0911SHGSANJP,15110JLBHYZLWEPL}, our scheme has stronger robustness against the temperature of environment.

The paper is organized as follows. In Sec. \uppercase\expandafter{\romannumeral 2}, we establish the model and present the equations of motion of the system. In Sec. \uppercase\expandafter{\romannumeral 3}, we give the steady-state mean values, linearize the quantum Langevin equations, and introduce the logarithmic negativity to quantify the entanglement between the oscillators. In Sec. \uppercase\expandafter{\romannumeral 4}, we numerically simulate the logarithmic negativity and show the modulation of entanglement. Finally we make a conclusion to summarize our results in Sec. \uppercase\expandafter{\romannumeral 5}.

\section{Model and equations of motion}\label{sec2}
The system considered consists of a Fabry-P\'{e}rot cavity containing one fixed partially transmitting mirror {\it A} and one movable totally reflecting mirror $B_1$ in contact with a thermal bath in equilibrium at temperature {\it T}, a charged oscillator $B_2$, and an OPA which is embedded into the cavity, as schematically shown in Fig.~1. The movable mirror $B_1$ can move along the cavity axis and is treated as a mechanical harmonic oscillator with effective mass $m$, frequency $\omega_{m_1}$, and energy decay rate $\gamma_{m_1}$ and is charged by the bias gate voltage $U_1$. The cavity mode couples to the mechanical oscillator $B_1$ via radiation pressure caused by the intracavity photons exerting on the movable mirror, while $B_1$ and $B_2$ are coupled by the Coulomb force~\cite{0572WKHDWUHSGKSCMGJMPRA,1286JQZYLMFYXPRA,1490PCMJQZYXMFZMZPRA,1591RXCLTSSBZPRA}. The cavity is coherently driven by an external laser with frequency $\omega_L$ and amplitude {\it E} from left side. The Hamiltonian of the system is given by
\begin{eqnarray}\label{e1}
H&=&\hbar\omega_cc^{\dag}c+\frac{\hbar\omega_1}{2}(p_1^2+q_1^2)+\frac{\hbar\omega_2}{2}(p_2^2+q_2^2)
+i\hbar E(c^{\dag}e^{-i\omega_Lt}-ce^{i\omega_Lt})\cr\cr
&&-\hbar G_0c^{\dag}cq_1+i\hbar C_g(e^{i\theta}c^{\dag2}e^{-2i\omega_Lt}-e^{-i\theta}c^{2}e^{2i\omega_Lt})+\frac{-k_eQ_1Q_2}{|d_0+q_1-q_2|},
\end{eqnarray}
where the first term is the free Hamiltonian for the cavity field with resonance frequency $\omega_c$ and annihilation (creation) operator $c$ ($c^{\dag}$). The second and third terms describe the vibration of the mechanical oscillators $B_1$ and $B_2$, respectively, and position operator $q_j$ and momentum operator $p_j$ satisfy the commutation relation $[q_j, p_j]=i$ ($j=1,2$). The fourth term is the pumping interaction between the cavity field and external deriving laser with $E=\sqrt{2\kappa P/\hbar\omega_L}$, where $P$ is the power of the driving laser and $\kappa$ is the cavity decay rate. The fifth term describes the optomechanical interaction between the cavity field and the mechanical oscillator $B_1$ with the optomechanical coupling strength $G_0=(\omega_c/L)\sqrt{\hbar/m\omega_m}$, where {\it L} is the separation between the mirror {\it A} and oscillator $B_1$ in the absence of radiation pressure and Coulomb interactions. The sixth term represents the coupling between the OPA and the cavity field, where $C_g$ is the nonlinear gain of the OPA and $\theta$ is the phase of the pump driving the OPA. The last term represents the Coulomb interaction of the two charged mechanical oscillators $B_1$ and $B_2$. $k_e$ denotes the electrostatic constant. $Q_j=C_jU_j$ is the charge carried by the electrode on oscillator $B_j$, where $C_j$ is the capacitance of the bias gate on $B_j$.  $d_0$ is the equilibrium separation between $B_1$ and $B_2$ in the absence of optomechanical and Coulomb interactions and $q_j$ represents the small deviation of $B_j$ from its equilibrium position due to the optomechanical and Coulomb interactions.
\begin{figure}\label{fig1}
\centering
\includegraphics[width=4.5in, height=2.78in]{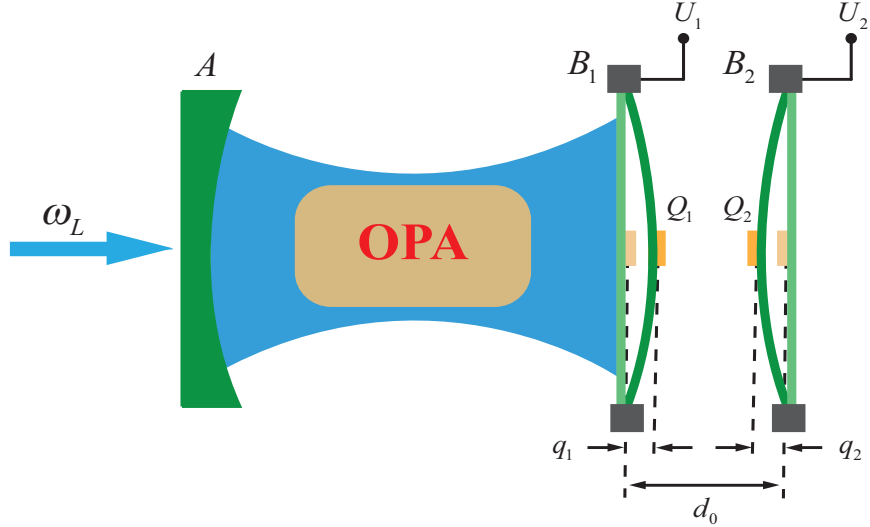}
\caption{(Color online) Schematic diagram of the system. The cavity optomechanical system consists of a fixed mirror $A$ and a mechanical oscillator $B_1$ which is coupled to the other mechanical oscillator $B_2$ under the action of the Coulomb interaction. An OPA is placed inside the cavity, and the pump of the OPA is not shown. The cavity is driven by the driving field $E$ with frequency $\omega_L$. The electrode carrying charge $Q_1$ ($Q_2$) on $B_1$ ($B_2$) is charged by the bias gate voltage $U_1$ ($U_2$). $d_0$ is equilibrium separation of $B_1$ and $B_2$. $q_1$ and $q_2$ are the small deviation of $B_1$ and $B_2$ from their equilibrium positions, respectively, due to the radiation pressure interaction and Coulomb interaction.}
\end{figure}

Since the mechanical deviation $q_j$ is comparatively small compared to the equilibrium separation $d_0$, i.e., $q_j \ll d_0$, the term of Coulomb interaction can be expanded to second-order of $(q_1-q_2)/d_0$ as follows
\begin{eqnarray}\label{e2}
H_{\mathrm{CI}}=\frac{-k_eC_1U_1C_2U_2}{|d_0+q_1-q_2|}
=\frac{-k_eC_1U_1C_2U_2}{d_0}\left[1-\frac{q_1-q_2}{d_0}+\left(\frac{q_1-q_2}{d_0}\right)^2\right],
\end{eqnarray}
where the linear term can be neglected via redefining the equilibrium positions of mechanical oscillators and the quadratic term includes a renormalization of the mechanical frequencies for both $B_1$ and $B_2$, whose effect is just a small frequency shift related to original frequencies, so it can be also neglected. Through further discarding the constant term, the Coulomb interaction can be reduced to the simpler form
\begin{eqnarray}\label{e3}
H_{\mathrm{CI}}=\hbar\lambda q_1q_2,
\end{eqnarray}
where $\lambda=2k_eC_1U_1C_2U_2/\hbar d_{0}^{3}$~\cite{0572WKHDWUHSGKSCMGJMPRA,1490PCMJQZYXMFZMZPRA,1591RXCLTSSBZPRA,0493LTPZPRL}.
In the interaction picture with respect to $\hbar\omega_L c^{\dag}c$, the system Hamiltonian can be rewritten as
\begin{eqnarray}\label{e4}
H&=&\hbar(\omega_c-\omega_L)c^{\dag}c+\frac{\hbar\omega_1}{2}(p_1^2+q_1^2)+\frac{\hbar\omega_2}{2}(p_2^2+q_2^2)+\hbar\lambda q_1q_2\cr\cr
&&+i\hbar E(c^{\dag}-c)-\hbar G_0c^{\dag}cq_1+i\hbar C_g(e^{i\theta}c^{\dag2}-e^{-i\theta}c^2).
\end{eqnarray}

A proper analysis of the system must consider the photon losses from the cavity and the Brownian noise from the environment. This can be accomplished via the dynamics of the system governed by Eq.~({\ref{e4}}) using quantum Langevin equation
\begin{eqnarray}\label{e5}
\dot{q_1}&=&\omega_1p_1,\cr\cr
\dot{p_1}&=&-\omega_1q_1-\gamma_{m1}p_1+G_0c^{\dag}c-\lambda q_2+\xi_1,\cr\cr
\dot{q_2}&=&\omega_2p_2,\cr\cr
\dot{p_2}&=&-\omega_2q_2-\gamma_{m2}p_2-\lambda q_1+\xi_2,\cr\cr
\dot{c}&=&-\left[\kappa+i(\omega_c-\omega_L)\right]c+iG_0cq_1+E+2C_ge^{i\theta}c^{\dag}+\sqrt{2\kappa}c_{in},
\end{eqnarray}
where $\gamma_{m2}$ is the damping rate for the oscillator $B_2$. $c_{in}$ is the input vacuum noise operator with zero mean value and nonzero correlation function $\langle c_{in}(t)c_{in}^{\dag}(t^{\prime})\rangle=\delta(t-t^{\prime})$~\cite{0368JZKPSLBPRA,0798DVSGAFHRBPTAGVVAZMAPRL,0877CGDVPTPRA}. The quantum Brownian noise $\xi_1$ ($\xi_2$) arises from the coupling between {\it $B_1$} ({\it $B_2$}) and its environment with zero mean value and correlation function~\cite{0163VGDVPRA}
\begin{eqnarray}\label{e6}
\langle\xi_j(t)\xi_j(t^{\prime})\rangle=\frac{\gamma_{mj}}{\omega_{mj}}\int\frac{\omega}{2\pi}e^{-i\omega(t-t^{\prime})}
\left[\coth\left(\frac{\hbar\omega}{2k_BT}\right)+1\right]d\omega,
\end{eqnarray}
where $k_B$ is the Boltzmann constant and {\it T} is the temperature of the environment in contact with the oscillators. However, quantum effects are revealed just for the oscillators with a large quality factor, i.e., $Q \gg 1$. In this limit, Eq.~({\ref{e6}}) can be further simplified to delta-correlated~\cite{0163VGDVPRA}
\begin{eqnarray}\label{e7}
\langle\xi_j(t)\xi_j(t^{\prime})+\xi_j(t^{\prime})\xi_j(t)\rangle/2\simeq\gamma_{mj}(2\bar{n}+1)\delta(t-t^{\prime}),
\end{eqnarray}
where $\bar{n}=(\mathrm{exp}\left\{\hbar\omega_{mj}/k_BT\right\}-1)^{-1}$ is the mean thermal excitation number. In the following we discuss the oscillator-oscillator entanglement in the regime where the system is stable.

\section{The oscillator-oscillator steady state entanglement}\label{sec3}
The stability of the steady state of the system is determined by a linearized analysis for small perturbation around the steady state~\cite{94DFWGJMQO}. We now first linearize the dynamics of the system. The nonlinear quantum Langevin equations can be linearized via rewriting each Hersenberg operator as its steady state mean-value plus an additional fluctuation operator with zero-mean value, i.e., $q_j=q_{js}+\delta q_j$, $p_j=p_{js}+\delta p_j$, and $c=c_j+\delta c$~\cite{1486MATJKFMRMP}. After inserting these expressions into the Langevin equations of Eq.~({\ref{e5}}), we can obtain a set of nonlinear algebraic equations for the steady state values and a set of quantum Langevin equations for the fluctuation operators~\cite{9449CFMPSBAHEGSRPRA}. Through setting all the time derivatives in algebra equations for the steady state value to zero, the steady state mean values of system are given by
\begin{eqnarray}\label{e8}
p_{1s}&=&0,\cr\cr
q_{1s}&=&\frac{G_0|c_s|^2}{\omega_{m1}-\frac{\lambda^2}{\omega_{m2}}},\cr\cr
p_{2s}&=&0,\cr\cr
q_{2s}&=&\frac{-\lambda}{\omega_{m2}}q_{1s},\cr\cr
c_{s}&=&\frac{\kappa-i\Delta+2C_ge^{i\theta}}{\kappa^2+\Delta^2-4C_g^2}E,
\end{eqnarray}
where $\Delta=\omega_c-\omega_L-G_0q_{1s}$ is the effective cavity detuning from the frequency of the input laser in the presence of the radiation pressure. The modification of the detuning by the $G_0q_{1s}$ term depends on the mechanical motion. The $q_{1s}$ represents the new equilibrium position of the oscillator $B_1$ relative to that in the absence of the optomechanical and Coulomb interactions and $c_s$ denotes the steady state amplitude of the cavity field.

In order to analyze the oscillator-oscillator steady state entanglement, we need to find out the fluctuations in the oscillators' amplitudes. So we are interested in the dynamics of small fluctuations around the steady state of the system. For generating the entanglement, generally, the cavity is intensively driven with a very large input power {\it P}, which means that at the steady state, the intracavity field has a large amplitude, i.e., $|c_s|\gg1$. In this strong driving limit, we can ignore some small quantities and get the linearized Langevin equations
\begin{eqnarray}\label{e9}
\delta\dot{q_1}&=&\omega_{m1}\delta p_1,\cr\cr
\delta\dot{p_1}&=&-\omega_{m1}\delta q_1-\gamma_{m1}\delta p_1-\lambda\delta q_2+G_0(c_s^{\ast}\delta c+c_s\delta c^{\dag})+\xi_1,\cr\cr
\delta\dot{q_2}&=&\omega_{m2}\delta p_2,\cr\cr
\delta\dot{p_2}&=&-\lambda\delta q_1-\omega_{m2}\delta q_2-\gamma_{m2}\delta p_2+\xi_2,\cr\cr
\delta\dot{c}&=&-\left(\kappa+i\Delta\right)\delta c+iG_0c_s\delta q_1+2C_ge^{i\theta}\delta c^{\dag}+\sqrt{2\kappa}c_{in}.
\end{eqnarray}

If we choose the phase reference of the cavity field so that $c_s$ is real and introduce the amplitude and phase fluctuations of the cavity field as $\delta X=(\delta c+\delta c^{\dag})/\sqrt{2}$ and $\delta Y=(\delta c-\delta c^{\dag})/\sqrt{2}i$, and the position and momentum fluctuations of the thermal noise as $X^{in}=(c_{in}+c_{in}^{\dag})/\sqrt{2}$ and $Y^{in}=(c_{in}-c_{in}^{\dag})/\sqrt{2}i$, Eq.~({\ref{e9}}) can be written as the matrix form
\begin{eqnarray}\label{e10}
\dot{f}(t)=Mf(t)+\eta(t),
\end{eqnarray}
where $f(t)$ is the column vector of the fluctuations and $\eta(t)$ is the column vector of the noise sources. Their transposes are
\begin{eqnarray}\label{e11}
f(t)^T&=&(\delta q_1, \delta p_1, \delta q_2, \delta p_2, \delta X, \delta Y),\cr\cr
\eta(t)^T&=&(0, \xi_1, 0, \xi_2, \sqrt{2\kappa}X^{in}, \sqrt{2\kappa}Y^{in});
\end{eqnarray}
and the matrix {\it M} is given by
\begin{eqnarray}\label{e12}
M=
\left[
\begin{array}{cccccc}
0~&\omega_{m1}~&0~&0~&0~&0\\
-\omega_{m1}~&-\gamma_{m1}~&-\lambda~&0~&G_m~&0\\
0~&0~&0~&\omega_{m2}~&0~&0\\
-\lambda~&0~&-\omega_{m2}~&-\gamma_{m2}~&0~&0\\
0~&0~&0~&0~&2C_g\cos\theta-\kappa~&2C_g\sin\theta+\Delta\\
G_m~&0~&0~&0~&2C_g\sin\theta-\Delta~&-(2C_g\cos\theta+\kappa)
\end{array}
\right],
\end{eqnarray}
where $G_m=\sqrt{2}G_0c_s$ is the effective optomechanical coupling. Remarkably, the quantum fluctuations of the field and the oscillator are now coupled by the much large effective coupling.

The solutions to Eq.~({\ref{e10}}) are stable only if all the eigenvalues of the matrix {\it M} have negative real parts. The stability conditions can be derived by applying the Routh-Hurwitz criterion~\cite{64AH,8735EXDCKPRA}, yielding the constrain conditions on the system parameters. Due to their expressions are considerable tedious, we don't report them here. However, we will satisfy the stability conditions of the system in the following analysis.

The solution of the first-order linear inhomogeneous differential Eq.({\ref{e10}}) can be solved as following form
\begin{eqnarray}\label{e13}
f(t)=u(t)f(0)+\int_0^tu(\tau)\eta(t-\tau)d\tau,
\end{eqnarray}
where the matrix $u(t)=\mathrm{exp}(Mt)$ and the initial condition $u(0)=I$ ({\it I} is the identity matrix).

An important type of continuous variable quantum states is the Gaussian states, which play a significant role in the foundation of quantum theory and also have potential applications in their relevant experiment~\cite{1284CWSPRGPNJCTCRJHSSLRMP}. The linearized effective Hamiltonian which corresponds to the linearized Langevin Eq.~(\ref{e9}) ensures that when the system is stable, it always reaches a Gaussian state whose information-related properties, such as entanglement and entropy, can be completely described by the symmetric $6\times6$ covariance matrix $V$~\cite{1284CWSPRGPNJCTCRJHSSLRMP,0740GAFIJPA} with components defined as
\begin{eqnarray}\label{e14}
V_{i,j}=\langle f_i f_j+f_j f_i\rangle/2.
\end{eqnarray}
From Eqs.~(\ref{e10}) and (\ref{e14}), we can derive a linear differential equation for the covariance matrix
\begin{eqnarray}\label{e15}
\dot{V}=MV+VM^T+D,
\end{eqnarray}
where $D$ is a diffusion matrix whose components are associated with the noise correlation function Eq.~(\ref{e7})
\begin{eqnarray}\label{e16}
D_{i,j}\delta(t-t^{\prime})=\langle \eta_i(t)\eta_j(t^{\prime})+\eta_j(t^{\prime})\eta_i(t)\rangle/2.
\end{eqnarray}
It is easy to obtain that $D$ is diagonal $D=\mathrm{diag}[0, \gamma_{m1}(2\bar{n}+1), 0, \gamma_{m2}(2\bar{n}+1), \kappa,\kappa]$. From the point of view of describing the dynamics of the system Gaussian states, Eq.~(\ref{e15}) is equivalent to the quantum Langevin equations Eq.~(\ref{e9}) but is more convenient for studying entanglement evolution.

The reduced $4\times4$ covariance matrix $\widetilde{V}$ for the mechanical oscillators $B_1$ and $B_2$ of interest here can be extracted from the full $6\times6$ covariance matrix $V$. If the reduced covariance matrix $\widetilde{V}$ is written as the block form
\begin{eqnarray}\label{e17}
\widetilde{V}=
\left[
\begin{array}{cc}
\Phi_1~&~\Phi_3\\
\Phi_3^T~&~\Phi_2
\end{array}
\right],
\end{eqnarray}
where $\Phi_k (k=1, 2,3)$ are $2\times2$ block matrices, then the entanglement of the two separated mechanical oscillators $B_1$ and $B_2$ quantified by the logarithmic negativity can be readily calculated~\cite{0265GVRFWPRA,0470GAASFI,0595MBPPRL}
\begin{eqnarray}\label{e18}
E_N=\mathrm{max}[0, -\ln(2\varrho)],
\end{eqnarray}
where $\varrho\equiv2^{-1/2}\left\{\Sigma(V)-\left[\Sigma(V)^2-4 \mathrm{det}V\right]^{1/2}\right\}^{1/2}$, with $\Sigma(V)\equiv \mathrm{det}\Phi_1 + \mathrm{det}\Phi_2 - 2 \mathrm{det}\Phi_3$. Therefore, a Gaussian state is entangled if and only if $\varrho<1/2$, which is equivalent to Simon's necessary and sufficient entanglement nonpositive partial transpose criterion for Gaussian states~\cite{0084RSPRL}.

\section{The modulation of oscillator-oscillator steady state entanglement under the action of the OPA}\label{sec4}
\begin{figure}\label{fig2}
\centering
\includegraphics[scale=0.5]{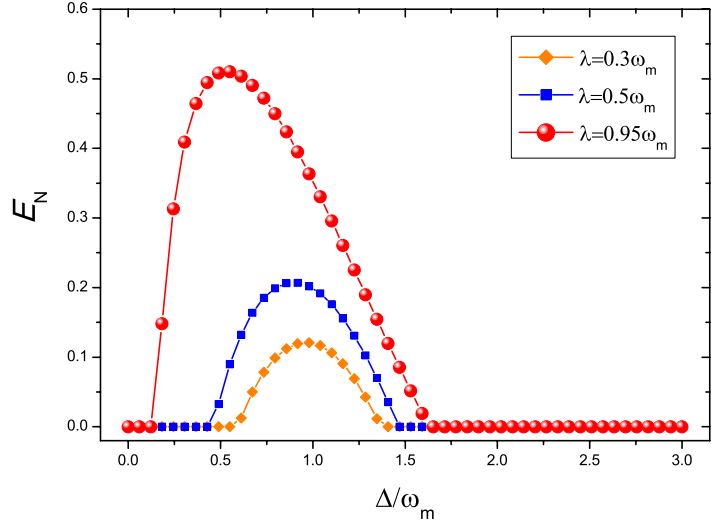}
\caption{(Color online) Plot of the logarithmic negativity $E_N$ as a function of the normalized detuning $\Delta/\omega_m$ for three different values of the Coulomb coupling strength $\lambda=0.3 \omega_m$ (orange diamond line), $\lambda=0.5 \omega_m$ (blue square line), and $\lambda=0.95 \omega_m$ (red sphere line) in the absence of the OPA ($C_g=0$). Here we have set the temperature of environment $T=4$ mK and driving power $P=50$ mW.}
\end{figure}
In this section, we numerically evaluate the logarithmic negativity $E_N$ between the two separated oscillators $B_1$ and $B_2$ to show the modulation of entanglement under the action of the OPA in experimentally accessible parameter regimes. Without loss of generality, we assume that all the parameters of the two mechanical oscillators to be the same, i.e., $\omega_{m1}=\omega_{m2}=\omega_m$, $\gamma_{m1}=\gamma_{m2}=\gamma_m$. We choose the parameters in our numerical calculations are based on the experiment conditions~\cite{084ASRRGAOATJKNP,06444SGHRBMPFBGLJBHKCSDBMAAZNature}: $\omega_m=200\pi$ MHz, $\gamma_m=200\pi$ Hz, $\kappa=88.1$ MHz, $m=5$ ng, $L=1$ mm, and the wavelength of driving laser $\lambda_0=810$ nm.

First, we illustrate the effect of the Coulomb interaction on the entanglement between the two separated mechanical oscillators in the absence of the OPA ($C_g=0$). The logarithmic negativity $E_N$ as a function of the normalized detuning $\Delta/\omega_m$ for three different values of the Coulomb coupling strength $\lambda=0.3 \omega_m$ (orange diamond line), $\lambda=0.5 \omega_m$ (blue square line), and $\lambda=0.95 \omega_m$ (red sphere line) at temperature $T=4$ mK and driving power $P=50$ mW in the absence of the OPA is shown in Fig.~2. As illustrated in previous section, as long as the logarithmic negativity $E_N > 0$, there is an entanglement between the oscillators, meaning that there is a quantum correlation between them, even though they are separated in space. From Fig.~2, one can clearly see that the larger the coupling parameter $\lambda$ is, the stronger the oscillators entangle and the broader the range of the entanglement is. The numerical result shows that if there is no the Coulomb coupling, it is not possible to entangle the oscillators which are separated in space. So the Coulomb interaction between the oscillators is the essential reason of the entanglement.

In the previous schemes, Ref.~\cite{0911SHGSANJP} proposed a scheme for entangling two separated oscillators by injecting squeezed vacuum light and laser light into the ring cavity. The entanglement between the oscillators can be modulated via the squeezing parameter of the input light. When the squeezed vacuum light is replaced by an ordinary vacuum light, i.e., the squeezing parameter of the input light is 0, there is no entanglement between the oscillators. However, on squeezing the injected vacuum light, the entanglement between the oscillators is emerged. When the squeezing parameter of the input light $r \in (0, 1)$, the entanglement becomes more and more stronger with the increase of $r$, while $r \in (1, 2)$, the entanglement becomes more and more weaker with the increase of $r$. So in this scheme modulation of the entanglement between the separated oscillators can be achieved by means of the squeezing parameter of the input light. Ref.~\cite{15110JLBHYZLWEPL} also proposed a method to coherently control the entanglement between two movable mirrors via placing the Kerr-down-conversion crystal consisting of Kerr nonlinear medium and OPA inside an optomechanical cavity. By the aid of the input squeezed vacuum field, the Kerr nonlinear medium can lead to stronger entanglement between the two movable mirrors and extend to wider entanglement region. Whereas the effect of the OPA on entanglement is completely opposite, it leads weaker entanglement and narrower entanglement region. So modulation of the entanglement between two separated movable mirrors can be achieved via the Kerr-down-conversion crystal. The above two schemes have the common point that they all resort to the external squeezed vacuum filed. Aa a matter of fact, the nonlinear interaction processes between light and OPA have been considered as important sources of squeezed state of the radiation field~\cite{97MOSMSZQO,94DFWGJMQO}. Next we take advantage of the nonlinearity of the OPA to generate the squeezed photons inside the optomechanical cavity which can interact directly with the oscillator $B_1$ to modulate the entanglement between the separated oscillators instead of resorting to additional external squeezed vacuum field.

\begin{figure}\label{fig3}
\centering
\includegraphics[scale=0.5]{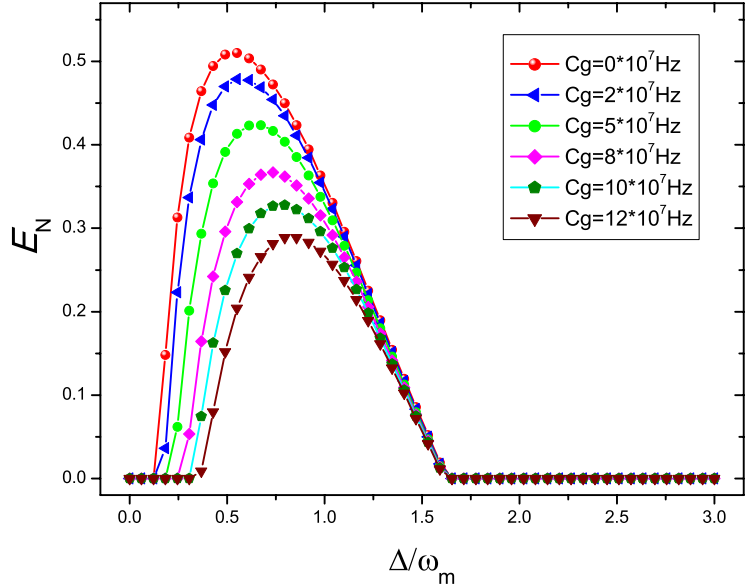}
\caption{(Color online) Plot of the logarithmic negativity $E_N$ as a function of the normalized detuning $\Delta/\omega_m$ for six different values of $C_g=0$ (red sphere line), $C_g=2\times10^7$ Hz (blue triangle line), $C_g=5\times10^7$ Hz (green circle line), $C_g=8\times10^7$ Hz (magenta diamond line), $C_g=10\times10^7$ Hz (olive pentagon line), and $C_g=12\times10^7$ Hz (wine triangle line). Here, we have set the Coulomb coupling strength $\lambda=0.95\omega_m$, the phase of the pump driving $\theta=0$, the temperature of environment $T=4$ mK, and the laser driving power $P=50$ mW.}
\end{figure}
We now show the effect of the gain of the OPA $C_g$ on the entanglement between the oscillators. We fix the Coulomb coupling strength $\lambda=0.95\omega_m$, the phase of the pump driving $\theta=0$, the temperature of environment $T=4$ mK, and the laser driving power $P=50$ mW. The logarithmic negativity $E_N$ as a function of the normalized detuning $\Delta/\omega_m$ for six different values of $C_g=0$ (red sphere line), $C_g=2\times10^7$ Hz (blue triangle line), $C_g=5\times10^7$ Hz (green circle line), $C_g=8\times10^7$ Hz (magenta diamond line), $C_g=10\times10^7$ Hz (olive pentagon line), and $C_g=12\times10^7$ Hz (wine triangle line) is shown in Fig.~3. From Fig.~3, we can find that the entanglement between the oscillators becomes more and more weaker and the entanglement region becomes more and more narrower with the increase of the gain of the OPA $C_g$ compared with the case of in the absence of OPA ($C_g=0$). Additionally, the position of the maximal entanglement moves to right with the increase of gain $C_g$ due to the fact that the injection of OPA strengths the steady intracavity field and in turn changes the effective deduning $\Delta$. This is very similar to such the case of the weaker Coulomb coupling strength in the absence of OPA. It is due to the fact that the squeezed photons generated by the OPA inside the optomechanical cavity lead to a stronger radiation pressure acting on the oscillator $B_1$ and there exists the competing effect between the radiation pressure interaction and the Coulomb interaction acting on the oscillator $B_1$.

\begin{figure}\label{fig4}
\centering
\includegraphics[scale=0.5]{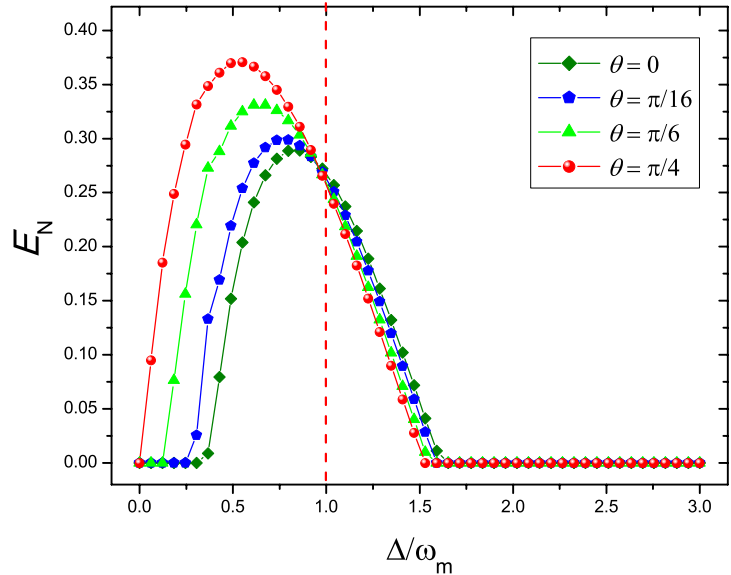}
\caption{(Color online) Plot of the logarithmic negativity $E_N$ as a function of the normalized detuning $\Delta/\omega_m$ for four different values of the phase of the pump driving the OPA $\theta=0$ (olive diamond line), $\theta=\pi/16$ (blue pentagon line), $\theta=\pi/6$ (green triangle line), and $\theta=\pi/4$ (red sphere line). Here, we have set the Coulomb coupling strength $\lambda=0.95\omega_m$, the gain of the OPA $C_g=12\times10^7$ Hz, the temperature of environment $T=4$ mK, and the laser driving power $P=50$ mW.}
\end{figure}
We next examine the effect of the phase of the pump driving the OPA $\theta$ on the entanglement between the oscillators. We fix the gain of the OPA $C_g=12\times10^7$ Hz and other parameters are as same as the Fig.~3. The logarithmic negativity $E_N$ as a function of the normalized detuning $\Delta/\omega_m$ for four different values of the phase of the pump driving the OPA $\theta=0$ (olive diamond line), $\theta=\pi/16$ (blue pentagon line), $\theta=\pi/6$ (green triangle line), and $\theta=\pi/4$ (red sphere line) is shown in Fig.~4. It can be clearly seen that the entanglement between the oscillators becomes more and more stronger with the increase of the phase $\theta$ when $\Delta < \omega_m$ for the fixed gain $C_g$ of the OPA. This is due to the fact that the degree of the squeezing of the squeezed photons generated by the OPA becomes more and more smaller with the increase of the phase $\theta$ for the fixed gain $C_g$ and the Coulomb interaction becomes the dominant factor compared with the radiation pressure interaction for the oscillator $B_1$. We can also find that when $\Delta=\omega_m$, all curves are intersected in one point. This can be interpreted as driving the system by the red-detuned laser $\Delta=\omega_m$ in the resolved sideband limit makes the optomechanical interaction between the cavity field and the oscillator $B_1$ like a beam-splitter interaction. In such case, the competing effect between the radiation pressure interaction and the Coulomb interaction acting on the oscillator $B_1$ maintains a balance.

\begin{figure}
\includegraphics[width=2.5in]{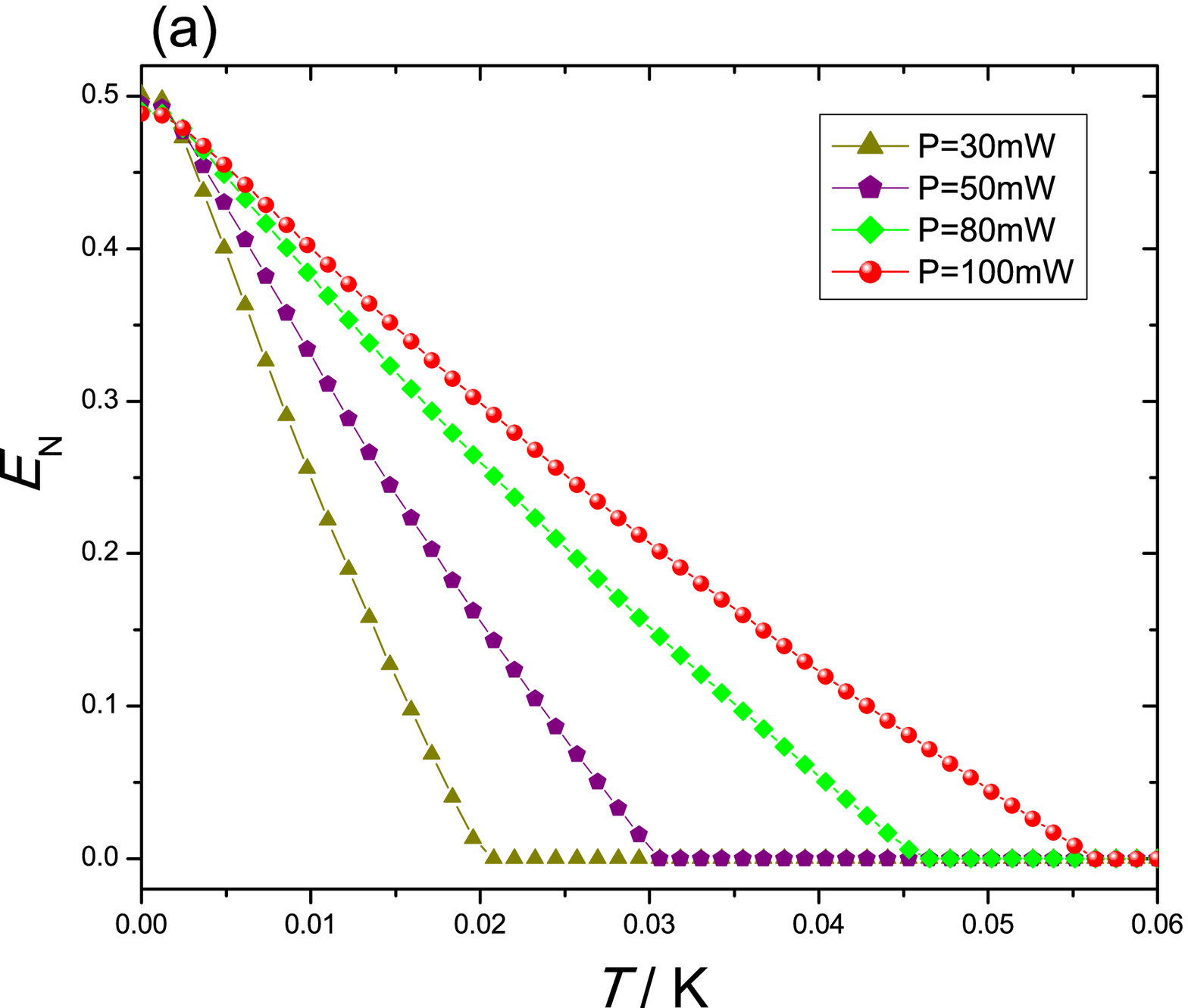}%
\hspace{0.3in}%
\includegraphics[width=2.5in]{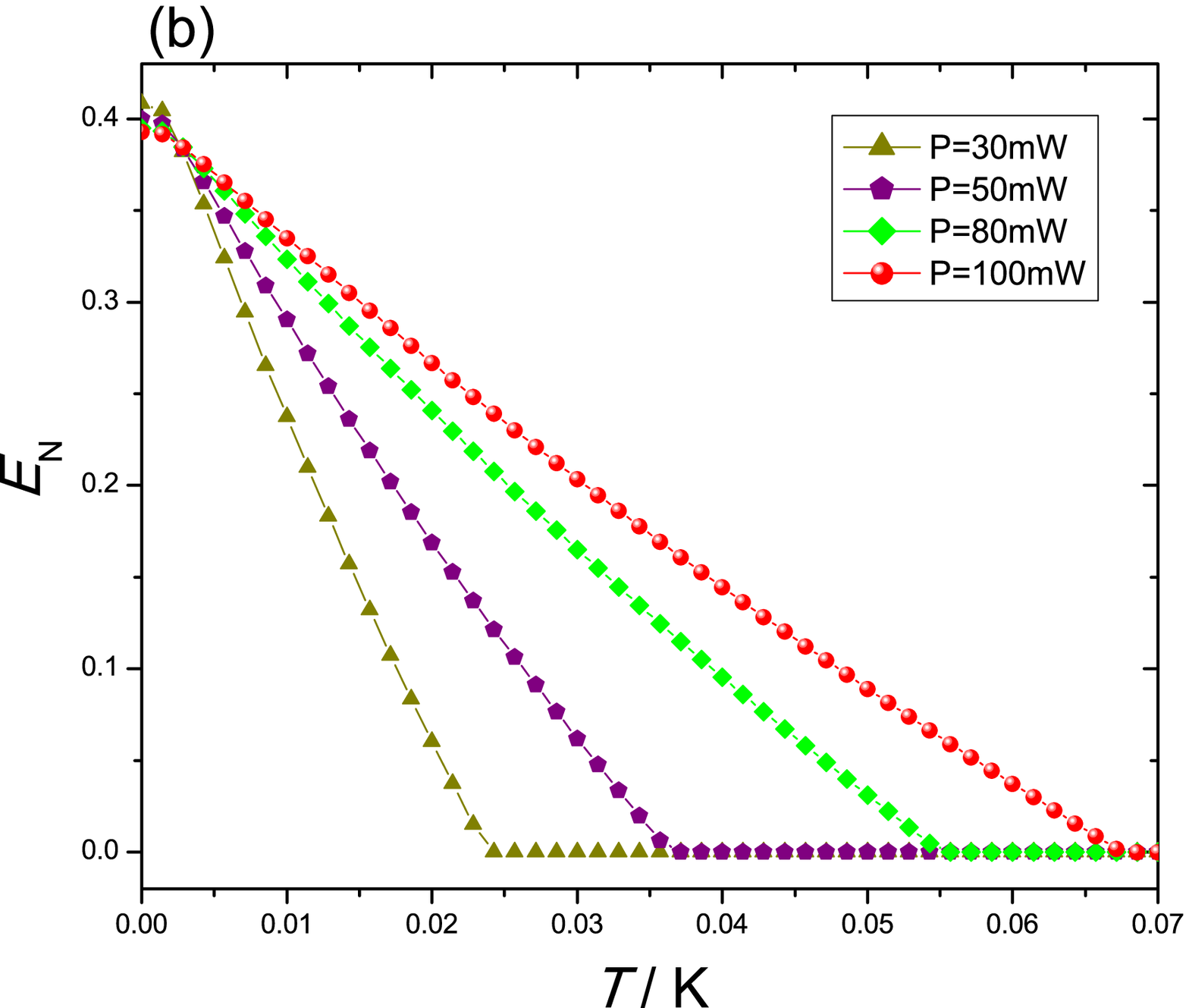}
\caption{(Color online) Plot of the logarithmic negativity $E_N$ as a function of the temperature for four different values of the laser power $P=30$ mW (dark yellow triangle line), $P=50$ mW (purple pentagon line), $P=80$ mW (green diamond line), and $P=100$ mW (red sphere line). Here, we have set $\lambda=0.95\omega_m$, $\Delta=0.75\omega_m$, $\theta=\pi/16$, and $C_g=2\times10^7$ Hz in (a), while $\lambda=0.95\omega_m$, $\Delta=0.75\omega_m$, $\theta=\pi/16$, and $C_g=8\times10^7$ Hz in (b).}
\end{figure}
In the following, we show the effect of the Brownian noise on the entanglement between the oscillators, i.e., the effect of the temperature of the environment. The logarithmic negativity $E_N$ as a function of the temperature for four different values of the laser power $P=30$ mW (dark yellow triangle line), $P=50$ mW (purple pentagon line), $P=80$ mW (green diamond line), and $P=100$ mW (red sphere line) when the Coulomb coupling strength $\lambda=0.95\omega_m$, $\Delta=0.75\omega_m$, and the phase of the pump driving $\theta=\pi/16$ is plotted in Fig.~5, wherein in Fig.~5(a) $C_g=2\times10^7$ Hz while $C_g=8\times10^7$ Hz in Fig.~5(b). It can be concluded that for the fixed gain $C_g$ of the OPA, as the temperature of the environment increases, the amount of entanglement monotonically decreases due to the thermal fluctuation which is as expected. The higher the temperature of the environment becomes, the stronger the thermal noise is. Then the entanglement between two oscillators is submerged by the strong thermal noise. The critical temperature of the entanglement is improved with the increase of the laser driving power for the fixed gain $C_g$ of the OPA and the numerical simulation results indicate that the robustness is obviously increased compared with previous schemes~\cite{0911SHGSANJP,15110JLBHYZLWEPL}. While for the fixed laser driving power, the critical temperature of the entanglement is higher with respect to the larger gain $C_g$ of the OPA. More importantly, it has been verified that the OPA inside a cavity can considerably improve the cooling of the oscillator by radiation pressure~\cite{0979SHGSAPRA}. So the OPA not only can improve the robustness of the entanglement, but also can cool the oscillator to its quantum ground state, which is very significant from the experimental point. Additionally, the relevant experimental investigation in such temperature requirement can be explored in the circuit cavity electromechanics~\cite{11471JDTDLMSAKCAJSJDWRWSNature}, which is easily cooled to temperatures below 100 mK.

Methods for detection of entanglement have been discussed in~\cite{0288SMVGDVPTPRL,0572MPADDVOATBAHEPL} and the entanglement properties between the oscillators can be verified by experimentally measuring the corresponding covariance matrix. It can be achieved by combining existing experimental techniques. The mechanical position and momentum can be measured  with the setup proposed in ~\cite{0798DVSGAFHRBPTAGVVAZMAPRL}, in which via adjusting the detuning and bandwidth of an additional adjacent cavity, both position and momentum of the oscillator can be measured by homodyning the output of the second cavity.

\section{Conclusions}\label{sec5}
In conclusion, we have shown that the modulation of entanglement between oscillators separated in space can be achieved via the squeezing cavity field generated by the OPA instead of directly injecting the squeezing field with the assistance of Coulomb interaction. We showed that the Coulomb interaction between the oscillators is the essential reason for the existence of entanglement. Through modulating the squeezing cavity field by the gain of OPA and the phase of the pump driving the OPA, the radiation pressure interaction between the cavity field and the oscillator obtains modulation accordingly. With the assistance of Coulomb interaction, we showed that under the action of competing effect between the radiation pressure interaction and Coulomb interaction, the entanglement between the oscillators can be modulated successfully. Moreover, in experimentally feasible regimes, the results of numerical simulation showed that the present scheme has stronger robustness against the temperature of environment compared with previous schemes.

\begin{center}
{\bf{ACKNOWLEDGMENTS}}
\end{center}

This work was supported by the National Natural Science Foundation of China under
Grant Nos. 11264042, 11465020, 61465013, 11564041, and the Project of Jilin Science and Technology Development for Leading Talent of Science and Technology Innovation in Middle and Young and Team Project under Grant No. 20160519022JH.

\end{document}